\documentclass{pic2012}
\def\runauthor{``Y.~Koshio et al''}
\def\shorttitle{Other neutrino oscillation measurements}
\usepackage{graphicx}
\begin{document}

\title{Other neutrino oscillation measurements}

\author{Yusuke Koshio}

\address{Kamioka Observatory, ICRR, University of Tokyo\\
456 Higashi-Mozumi, Kamioka, Hida, Gifu, 5061205, Japan\\
E-mail: koshio@icrr.u-tokyo.ac.jp }

\maketitle

\abstracts{
Neutrino oscillation results from several experiments and sources are discussed.
Recent results from solar neutrino measurements
by Super-Kamiokande and Borexino, atmospheric neutrino
measurements from Super-Kamiokande, and accelerator
neutrino measurements by MINOS and OPERA are the main topics
of this document.}

\section{Introduction}\label{sec:int}
 Neutrino oscillations were first discovered by the observation
of atmospheric neutrinos in Super-Kamiokande (SK) in 1998.
This discovery opened a new window to explore physics beyond
that predicted by the standard model of elementary particles.
The neutrino mixing matrix, which translates neutrino mass
eigenstates into flavor eigenstates, is usually parametrized
by three mixing angles ($\theta_{12},\theta_{23},\theta_{13}$),
and one CP-violating Dirac phase ($\delta$) in the three flavor neutrino
framework as \footnote{here, two Majorana CP phases ($\alpha_{21},
\alpha_{31}$) are omitted, since neutrino oscillation
does not depend on them.}
~\cite{3flavor-1}~\cite{3flavor-2}~\cite{3flavor-3}
\begin{eqnarray}
\left(
\begin{array}{c}
\nu_e\\
\nu_{\mu}\\
\nu_{\tau}
\end{array}
\right)
=
\left(
\begin{array}{ccc}
1 & 0 & 0\\
0 & c_{23} & s_{23} \\
0 & -s_{23} & c_{23}
\end{array}
\right)
\left(
\begin{array}{ccc}
c_{13} & 0 & s_{13} e^{-i\delta} \\
0 & 1 & 0\\
-s_{13} e^{-i\delta} & 0 & c_{13}
\end{array}
\right)
\left(
\begin{array}{ccc}
c_{12} & s_{12} & 0\\
-s_{12} & c_{12} & 0\\
0 & 0 & 1
\end{array}
\right)
\left(
\begin{array}{c}
\nu_1\\
\nu_2\\
\nu_3
\end{array}
\right),
\end{eqnarray}
where $c_{ij}$ and $s_{ij}$ represent $\cos\theta_{ij}$ and
$\sin\theta_{ij}$, respectively.
Neutrino oscillation frequencies are determined by the neutrino
mass differences, $\Delta m^2_{21}=m^2_2-m^2_1$ and
$\Delta m^2_{32}=m^2_3-m^2_2$, where $m_1$, $m_2$, and $m_3$
are the three mass eigenvalues.

 In 2012, substantial progress in neutrino physics was made 
with the discovery of $\theta_{13}$ by the reactor neutrino and T2K experiments.
\cite{reactor}\cite{t2k} : $\sin^22\theta_{13} = 0.099 \pm 0.014$.
This non-zero value makes the measurement of
the CP-violation in the lepton sector possible.
Among the oscillation parameters, $\theta_{12}$ and 
$\Delta m^2_{21}$ have been measured by solar and
reactor neutrino experiments, and $\theta_{23}$ and 
$\left|\Delta m^2_{32}\right|$ have been measured by atmospheric
and accelerator neutrino experiments. Recent results
from these measurements are presented in the following sections.

\section{Solar neutrinos}\label{sec:sol}
\begin{figure}[!thb]
\begin{center}
\includegraphics[width=10cm,bb=0 0 720 540]{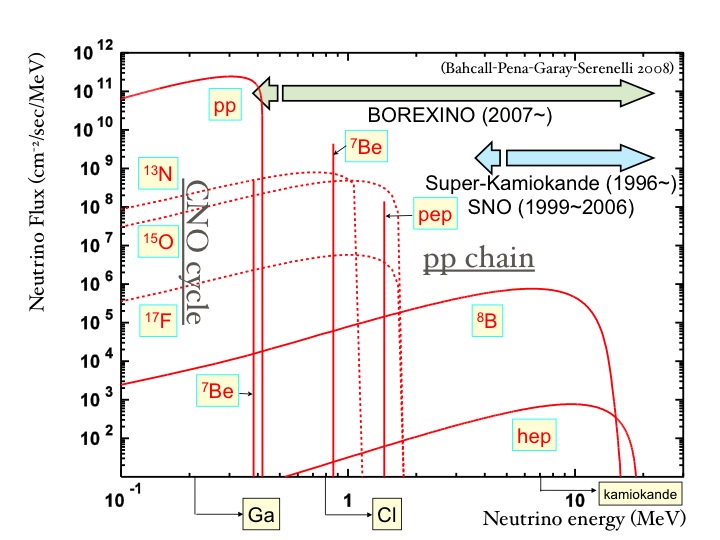}
\caption[*]{Solar neutrino spectrum expected by the Standard Solar Model. The observable energy region for each experiments are also overlaid.}
\label{fig:ssm}
\end{center}
\end{figure}

 The nuclear fusion reaction in main sequence stars like the Sun is as follows:
\begin{equation}
  4p \to \alpha + 2e^+ + 2\nu_e + 26.73\mbox{MeV}.
  \label{eq:nuc}
\end{equation}
Neutrinos generated in the core of the Sun reach its surface almost
immediately ($\sim$2~sec) and arrive at the earth $\sim$8~minutes later.
Therefore, solar neutrino measurements carry direct information about the current status of the center of the Sun.
Neutrinos are generated through the pp-chains and the CNO cycle.
The 'Standard Solar Model' (SSM)~\cite{bi:ssm}, which includes these reactions, 
predicts the solar neutrino flux and energy spectrum with good accuracy.
Fig.~\ref{fig:ssm} shows the predicted spectra and
observable energy region for several experiments.
Each of the modern solar neutrino experiments offers its 
own advantages and contributes to an overall complementarity 
among all measurements:  
SK collects a large statistical sample and makes a precise measurement of the energy spectrum,
SNO measured the solar neutrino flavors separately,
and BOREXINO makes measurements at low energies.
There are two primary motivations for solar neutrino research.
The first is the study of the energy generation mechanism of the sun,
and the second is the study of the neutrino.
The latter is the concentration of this paper.
\subsection{Borexino}\label{sec:solarbx}
 Borexino is an ultra-radiopure large volume liquid scintillator
detector, located underground (3500m water equivalent) at the Gran Sasso in Italy. 
The inner core scintillator serves as the target for neutrino detection and
consists of 270 tons of pseudocumene doped with 1.5 g/l of PPO. 
It is contained in a 4.25m radius spherical nylon vessel.
Scintillation light is detected by 2212 8-inch photomultiplier tubes
(PMTs) mounted on a stainless steel sphere (SSS).
In order to reduce external $\gamma$ and neutron backgrounds from PMTs and
the surrounding rock, the inner scintillator is shielded by 1000 tons of pseudocumene
doped with 5.0 g/l of dimethylphthalate (DMP) in a buffer region within the SSS,
and 2000 tons of pure water around it.
The external water tank is also used to detect residual cosmic ray muons
crossing the detector by observing their Cherenkov radiation.

 Solar neutrinos are detected via elastic scattering on electrons
in the liquid scintillator. The high light yield
($\sim$500 photoelectrons/MeV) of the scintillator enables 
a low energy threshold ($\sim$250keV) and good energy energy resolution
($\sim 5\%/\sqrt{\mathstrut E/(1MeV)}$).
Further, pulse shape discrimination between $\alpha$ and $\beta$ events is possible.
However, since there is no way to distinguish the neutrino signal from $\beta$-like events
induced by radioactive contaminants, extreme radiopurity is required.
Using a custom liquid scintillator purification system the 
$^{238}$U and $^{232}$Th contaminations has been reduced below the design value of 10$^{-16}$ g/g. 
This level of radiopurity is sufficient to measure not only $^7$Be solar neutrinos, 
but also $^8$B and potentially the pep and CNO neutrinos as well. 
More details are reported in~\cite{bi:purify}.

 The Borexino data set corresponds to a 740.66 day exposure taken from May 16, 2007
to May 8, 2010.~\cite{bi:bx2}
The fiducial exposure in this analysis is equivalent to 153.62 ton$\cdot$year.
The $^7$Be solar neutrino rate was evaluated using a spectral fit
and found to be $46.0\pm1.5(stat.)\pm1.3(sys.)$ counts/day/100ton. 
The total uncertainty is 4.7\%, with contributions from 
systematic errors at $+$3.4 and $-$3.6\%, 
which is an improvement over the previous result's $\sim$12\%.~\cite{bi:bx1}
Compared to the expected flux from the SSM, 67$\sim$74 counts/day/100ton,
the ``no oscillation'' hypothesis can be rejected and including 
oscillations the Borexino result is in good agreement with the MSW-LMA scenario.

 In the $^7$Be solar neutrino energy region, the difference in the day and night 
fluxes can be used to distinguish between the MSW-LMA and MSW-LOW models.
This difference is expected to be about 20\% for the the MSW-LOW model
with no difference for the MSW-LMA model. Looking at the day-night asymmetry
no significant effect can be seen in the Borexino data.
The data exclude the MSW-LOW region at more than 8~$\sigma$.~\cite{bi:bx3}

 The other recent result from Borexino is the first direct measurement of pep
neutrinos. Pep neutrino detection is not easy because there are unavoidable
cosmogenic backgrounds, such as the $\beta^+$ decay from $^{11}$C, in the signal region.
In 95\% of cases at least one free neutron is generaged in the $^{11}$C
production process. This background can be reduced by performing a space and
time veto using a three-fold coincidence between the parent muon, the neutron and
the subsequent 11C decay .
After this reduction, a spectral fit was applied, and a clear pep signal can
be seen. The observed pep solar neutrino rate is $3.1\pm0.6\pm0.3$
counts/day/100ton which is equivalent to $(1.6\pm0.3)\times10^8cm^{-2}s^{-1}$.
An upper limit on the CNO neutrino flux was also established 
at less than $7.7\times10^8cm^{-2}s^{-1}$ at 95\%C.L, which is the strongest
constraint to date.~\cite{bi:bx4}

\begin{figure}[!thb]
\begin{center}
\includegraphics[width=10cm,bb=0 0 720 540]{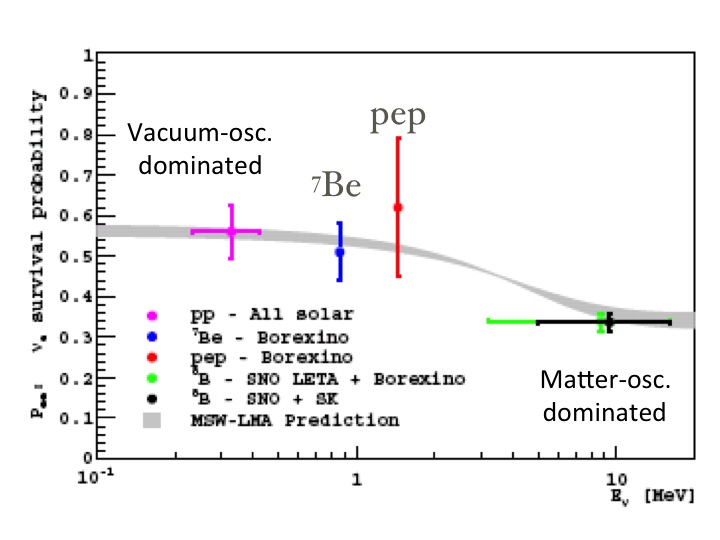}
\caption[*]{Observed electron neutrino survival probability as a function of energy by Borexino experiment~\protect\cite{bi:bx4}. The gray shape shows the expected from MSW-LMA neutrino oscillation scenario.}
\label{fig:bx}
\end{center}
\end{figure}

 Solar neutrino measurements at low energies with  Borexino can probe
both the vacuum and matter-enhanced neutrino oscillation regimes.
Figure~\ref{fig:bx} shows the neutrino survival probability as a function of energy.
The results of the $^7$Be and pep measurements from Borexino are in good agreement with
MSW-LMA scenario.
\subsection{Super-Kamiokande}\label{sec:solarsk}
 The Super-Kamiokande detector is a cylindrical tank (39.3~m in diameter and 41.4~m in height)
filled with 50~kilotons of pure water and lined with 11,146 20-inch PMTs.
It is located 1000~m underground (2,700~m water equivalent)
in the Kamioka mine in Gifu Prefecture, Japan.
Cherenkov light from charged particles are observed by the PMTs.
SK detects solar neutrinos through neutrino-electron elastic scattering, 
$\nu + e \rightarrow \nu + e$, where the energy, direction and time of the
recoil electron are measured.
Due to its large fiducial mass, 22.5 kiloton, SK gives the most precise
measurement of the solar neutrino flux and provides accurate information on the 
energy spectrum and its time variation. 

 By the end of March 2012, 1069.3 days of SK-IV solar neutrino data for
analysis was taken. One of the most important improvements to these measurements has been 
the reduction of the energy threshold, which now triggers at 100\% efficiency
at 4.0 MeV electron kinetic energy. A clear solar neutrino signal in the
3.5-4.0 MeV energy region can also be seen at more than $7~\sigma$.
During SK-IV the measured $^{8}$B flux is (2.34$\pm$0.03(stat.)$\pm$0.04(sys.)
$\times$ 10$^{6}$cm$^{-2}$s$^{-1}$),
which is consistent with previous measurements from SK-I, II, and III.~\cite{bi:sk1}\cite{bi:sk2}\cite{bi:sk3}

 Assuming the current measured values of the MSW-LMA oscillation parameters are correct,
a distortion of the solar neutrino energy spectrum is expected. Indeed, 
it is expected to increase with decreasing energy (the ``up-turn'' effect) 
in the sub-MeV energy region due to the transition from matter dominated oscillations 
to vacuum oscillations. 
In the latest SK results, a significant distortion cannot be seen.

 Concerning differences in the day and night fluxes the expected flux asymmetry,
defined as $A_{DN} = (day-night)/\frac{1}{2}(day+night)$, is about 2\% based on
the current understanding of neutrino oscillation parameters.
Although this is not a large effect, long term observations by SK
enable discussion of a finite value of the day-night asymmetry.
The $A_{DN}$ value using the combined SK-I to SK-IV data
is $-2.8\pm1.1\pm0.5\%$, which is a 2.3$\sigma$ difference from zero.
It is consistent with the expectation using the best fit $\Delta m^2$
from both KamLAND and the global solar analysis.

 A global solar neutrino oscillation analysis has been performed including
all SK data as well as the most recent results from other experiments.
This analysis was then compared and combined with the reactor neutrino results
from KamLAND~\cite{bi:kam}.

\begin{figure}[!thb]
\begin{center}
\hspace{-5mm} 
\includegraphics[width=9cm,bb=0 0 720 540]{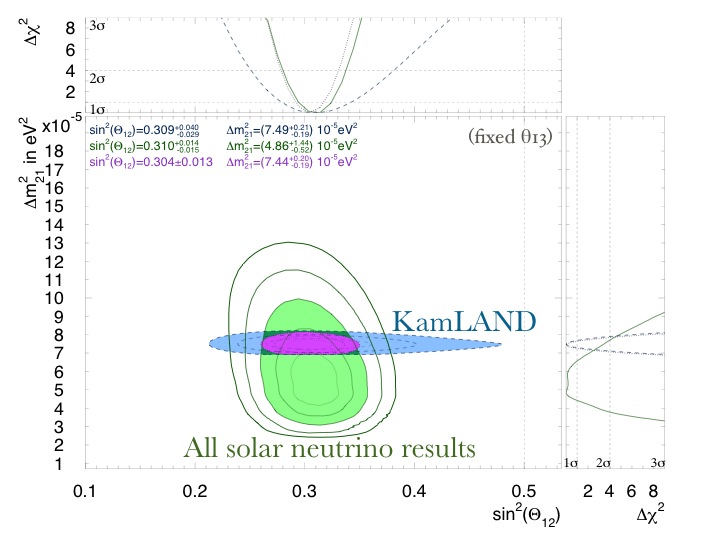}
\caption[*]{Neutrino oscillation parameter allowed regions in the $\Delta m^{2}_{21}$ and sin$^{2}\theta_{12}$ plane with $\sin^2\theta_{13}$ fixed at 0.025 from the global solar neutrino analysis (green) and the KamLAND reactor neutrino data (light blue)~\protect\cite{bi:sk4}. The purple area shows the combined contour of the global solar and the KamLAND reactor analysis. The curves are drawn for each 1-$\sigma$ step from 1-5~$\sigma$ for the global solar, and from 1-3~$\sigma$ for the KamLAND and solar+KamLAND results. Contours at 3~$\sigma$ are filled with their colors.}
\label{fig:sksol}
\end{center}
\end{figure}

 Figure \ref{fig:sksol} shows the allowed region of neutrino oscillation 
parameters in the $\Delta m^{2}_{21}$ and sin$^{2}\theta_{12}$
plane assuming $\sin^2\theta_{13}$ is fixed at 0.025.
The obtained parameters from the global solar analysis are
$\Delta m^{2}_{21}=(4.86^{+1.44}_{-0.52})\times 10^{-5}$eV$^{2}$ and
sin$^{2}\theta_{12} = 0.310^{+0.014}_{-0.015}$.
Comparing these values with those from KamLAND,
($\Delta m^{2}_{21}=(7.49^{+0.20}_{-0.19})\times 10^{-5}$eV$^{2}$ and
sin$^{2}\theta_{12} =0.309^{+0.039}_{-0.029}$), there is a 1.8~$\sigma$ 
tension in the $\Delta m^{2}_{21}$ results, which is evident in the figure.
Combining the global solar data with KamLAND, the oscillation parameters become
$\Delta m^{2}_{21}=(7.44^{+0.20}_{-0.19})\times 10^{-5}$eV$^{2}$ and
sin$^{2}\theta_{12} = 0.304\pm0.013$.

\section{Atmospheric neutrinos}\label{sec:atm}
Atmospheric neutrinos are produced in the decay chains of pions
and kaons generated by cosmic ray interactions
in the atmosphere. Both muon and electron neutrinos are 
produced and their predicted ratio has been calculated to an accuracy of better than 5\,\%.
The most important feature is that the fluxes of upward- and
downward-going neutrinos are expected to be nearly equal for
$E_\nu >$(a few GeV) where the effect of the geomagnetic field
on primary cosmic rays is negligible. 

In Super-Kamiokande atmospheric neutrino events are categorized by their topology and visible energy.
Fully contained (FC) events deposit all of their Cherenkov light
within the inner detector, while partially contained (PC) events have
exiting tracks which deposit energy in the outer detector.
The FC events are classified into ``sub-GeV'' ($E_{vis}<1330$~MeV) 
and ``multi-GeV'' ($E_{vis}>1330$\,MeV). 
These events are further separated into sub-samples based on the number
of observed Cherenkov rings. The single- and multi-ring samples are then divided into
electron-like (e-like) or muon-like ($\mu$-like) samples depending on  the pattern
identification of their most energetic Cherenkov ring. 
The sub-GeV samples are additionally divided based on their number of
decay-electrons and their likelihood of being a $\pi^0$.
The PC events are separated into "OD stopping" and "OD through-going"
categories based on the amount of light deposited in the OD by the exiting particle.
Energetic atmospheric $\nu_{\mu}$'s passing through the Earth interact
with rock surrounding the detector and produce leptons via charged current
interactions. When muons are produced with enough energy to traverse the 
rock and enter SK, they are observed as upward-going muon-like tracks.
These upward-going muons are classified into two types: 
``upward through-going muons,'' which have passed through
the detector, and ``upward stopping muons,''
which come into and stop inside the detector.
The upward through-going muons are subdivided into "showering" and
"non-showering" based on whether their Cherenkov pattern is
consistent with light emitted from an electromagnetic-magnetic shower produced
by a very high energy muon.

\begin{figure}[!thb]
\begin{center}
\includegraphics[width=7cm,bb=0 0 568 550]{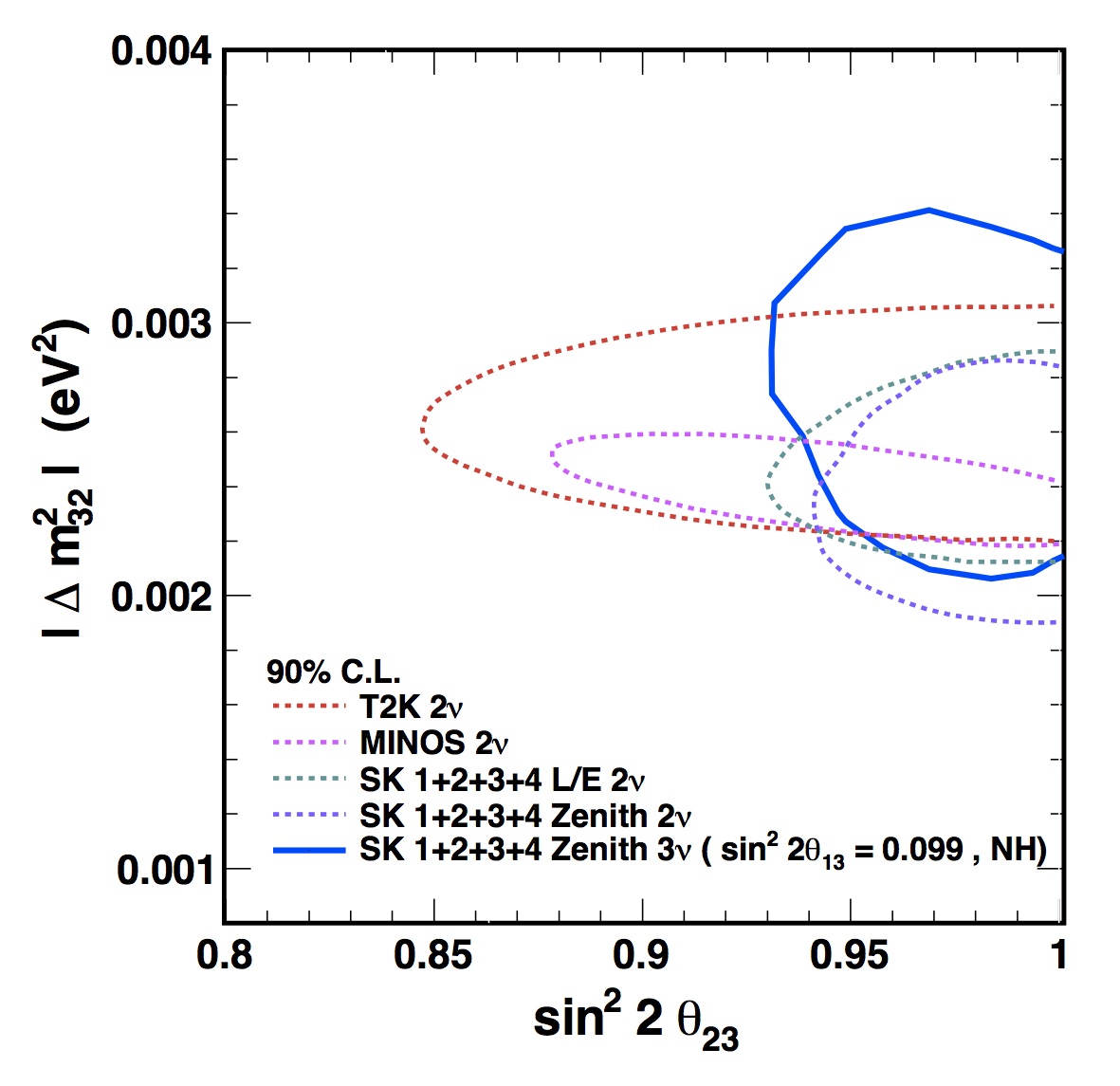}
\end{center}
\caption{Allowed regions for neutrino oscillations between the second and third generation observed in the SK atmospheric neutrino sample and overlaid with the results from other experiments~\protect\cite{bi:atm}.}
\label{fig:combined}
\end{figure}

 Super-Kamiokande has accumulated 3909 days of data of atmospheric neutrino data since it started in 1996.
At first, Super-K carried out a two-neutrino $\nu_{\mu} \rightarrow \nu_{\tau}$ oscillation analysis.
The purple dotted contour in Figure~\ref{fig:combined} shows
the allowed neutrino oscillation parameter regions from 
this analysis. The best fit oscillation
parameters are $\sin^2 2\theta=0.99$ and $\Delta m^2 = 2.30 \times
10^{-3}\rm eV^2$.

 In the standard two-neutrino framework, the survival $\nu_{\mu}$ probability 
is given by a sinusoidal function of $L/E$, where $L$ is the distance
traveled by the neutrino and $E$ is its neutrino energy.
Super-K has performed an oscillation analysis where the data 
have been binned in the combined variable $L/E$ to improve sensitivity to this sinusoidal dependence.
The allowed oscillation parameter region from this analysis is shown 
as the green dotted contour in Figure~\ref{fig:combined}.
This result is consistent with that from the 
analysis using the zenith angle binning described above.
The observed $L/E$ distribution gives evidence that the neutrino 
survival probability obeys the sinusoidal
function predicted by neutrino oscillations.
This result also excludes other models such as neutrino decay
(4.0$\sigma$) and neutrino decoherence
(4.8$\sigma$), which predict L/E spectra that lack the characteristic features 
of neutrino oscillations.

 The two-flavor neutrino oscillation model ignores the existence of a third neutrino 
mass eigenstate and in effect assumes that $\theta_{13}$=0
and that the mixing contribution from the solar parameters, $\Delta m^2_{12}$
and $\theta_{12}$, are negligible. However, the results of the solar 
and reactor neutrino experiments indicate that the solar parameters produce 
oscillation effects within the energy range relevant to atmospheric neutrinos.
Further a non-zero value for $\theta_{13}$ has recently been established.
In this scenario, the presence of matter along the path of neutrinos traveling 
upward through the earth induces a resonant enhancement of the 
$\nu_{\mu} \rightarrow \nu_{e}$ oscillation probability for a range of energies. 
This effect will manifest in the atmospheric neutrino data as an observable 
excess of upward-going multi-GeV electron neutrino events.
Additionally, the effects of the solar parameters
and non-maximal atmospheric mixing are observable as sub-leading oscillation effects
on the event rate of the Sub-GeV electron-like samples. 
If the CP violating term $\delta_{cp}$ is also considered, there are
additional sub-dominant oscillation effects predicted across many
of the SK atmospheric neutrino samples.
The blue contour in Figure~\ref{fig:combined} shows the allowed region
for the three flavor neutrino oscillation model 
assuming the neutrino mass hierarchy is normal. 

 Super-Kamiokande has also performed an extended oscillation analysis including
all the mixing parameters and the CP violating term.
The matter effect in the Earth is considered in this calculation
and both the normal and inverted mass hierarchies are tested.
The best fit to $\Delta m^2_{32}$ is $2.66\times 10^{-3} eV^2$
for both hierarchy assumptions. The allowed region for $\sin^2\theta_{23}$ at 90\% C.L. 
is $0.391 < \sin^2\theta_{23}<0.619$ for the normal hierarchy,
and $0.393 < \sin^2\theta_{23}<0.630$ for the inverted hierarchy.
All values of $\delta_{CP}$ are allowed at 90\% C.L, and no indication for either mass hierarchy is seen
in the data.
%

 Tau events, which are produced from $\nu_{\mu}$ that have oscillated into $\nu_{\tau}$ 
and subsequently undergo charged current interactions are expected to be observed in SK. 
Since there is no $\nu_{\tau}$ component in the primary atmospheric flux 
an observation of these events would be direct evidence for $\nu_{\mu}$ to $\nu_{\tau}$ oscillation.
However, the detection of $\nu_{\tau}$ events in SK is challenging.
Since the atmospheric neutrino flux falls as $1/E^{3}$ and $\nu_{\tau}$ charged current
interactions only ocurr above the $\tau$ production threshold, 3.5~GeV,
the expected rate at Super-K is only $\sim 1$/kton/year.
Further tau events are difficult to identify individually 
since they tend to produce multiple visible particles in SK.
An analysis is performed that employs a neural network technique to discriminate
events from the hadronic decays of $\tau$ in SK from atmospheric $\nu_e$ and
$\nu_{\mu}$ background events. Tau events are expected to appear as upward-going
events since this is the region of the observed $\nu_{\mu}$ to $\nu_{tau}$ oscillation effect.
The zenith angle distribution of the selected data are compared with the 
spectra produced from the sum of the expected backgrounds and tau signal. 
The background and signal normalizations are treated as free parameters and 
the bet fit signal excess corresponds to 180.1
$\pm$ 44.3 (stat) $^{+17.8}_{-15.2}$ (syst) $\nu_{t}$ events.
The significance of this observation is 3.8~$\sigma$.

\section{Other accelerator neutrino experiments}\label{sec:acc}
 The T2K experiment played an important role in the discovery of 
$\theta_{13}$, showing the first indication of electron neutrino appearance 
induced by this parameter. Other accelerator experiments have made several 
improved measurements of neutrino oscillation parameters. 
The results from MINOS and OPERA are described in this section.
\subsection{MINOS}\label{sec:minos}
 The MINOS experiment uses the NuMI neutrino beam from Fermilab,
which operates with 120 GeV/c protons directed onto a graphite
target.
The neutrino beam is derived from decay products of pions, kaons, and muons.
By changing the sign of the applied current to the magnetic focusing horns,
MINOS can focus either positively or negatively charged hadrons
and create either a neutrino or antineutrino beam, respectively.
The peak of the neutrino energy spectrum is around 3~GeV.
The total exposure of MINOS up until 2012 corresponds to $10.7\times 10^{20}$ 
protons on target (POT) for neutrinos and $3.4\times 10^{20}$ POT for antineutrinos.
MINOS uses two magnetized iron-scintillator detectors separated by a distance of 734km,
a 5.4~kton far detector and a 980~ton near detector.
Due to the magnetization, it is possible to detect neutrinos
and antineutrinos event-by-event based upon the curvature of out-going particles 
in the interaction. Both neutrino and antineutrino beam data
as well as atmospheric neutrino data have been studied at MINOS.

\begin{figure}[!thb]
\begin{center}
\includegraphics[width=7cm,bb=0 0 672 540]{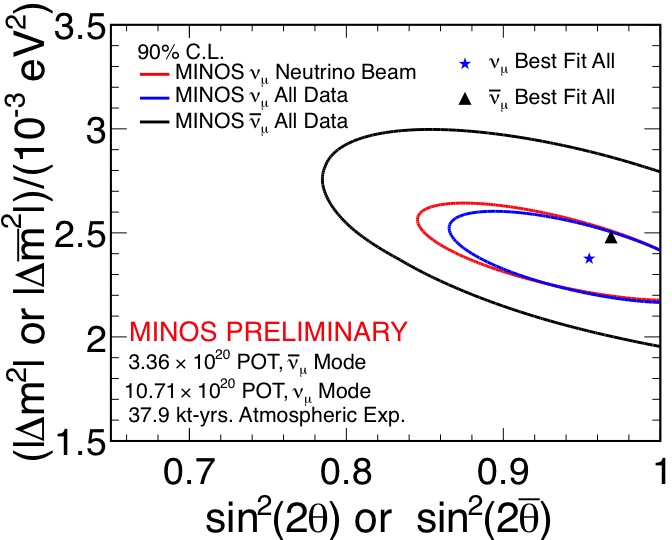}
\end{center}
\caption{Allowed neutrino oscillation parameter regions from the MINOS data~\protect\cite{minos2}.}
\label{fig:minos2}
\end{figure}

Figure~\ref{fig:minos2} shows the neutrino oscillation allowed region
extracted from all of the latest MINOS data~\cite{minos2}.
The parameter regions allowed by the neutrino and antineutrino data are consistent,
and they are consistent with the Super-Kamiokande and T2K measurements.
%
%
%
\subsection{OPERA}\label{sec:opera}
 The main purpose of the OPERA experiment is the observation of $\nu_{\tau}$
appearance from neutrino oscillations. The principle is same as SK, and 
relies on the identification of charged current $\nu_{\tau}$ interactions. 
However, unlike SK, OPERA has the ability to identify 
tau events on an individual basis. 

 In this experiment, the CNGS (CERN Neutrinos to Gran Sasso) beam is used.
It is a dedicated wide band neutrino beam with a mean energy of 17~GeV.
Neutrinos are generated from the decays of several particles, which are produced
in the interaction of the 400~GeV proton beam (SPS) with the target and 
focused by magnetic focusing horns. 
Muon neutrinos represent the majority of the flux and  $14.2\times 10^{19}$ POT 
have been collected up until 2011.
The detector is located in the underground Gran Sasso laboratory
730km away from the beam generation point. 
OPERA's measurements are based on the direct
observation of the $\tau$ decay topology using nuclear emulsion films,
which provide reconstruction accuracies of order 1~$\mu$m in position
and 1~mrad in angle. Since the neutrino interaction rate is small
a large target mass is also needed. The basic target unit is a brick of 10.2 x
12.8 x 7.9~$cm^{3}$ made of 56 lead plates and 57 emulsion films.
There are altogether 150,000 bricks in OPERA for a total mass of 1.25 kiloton.
Though the analysis is still ongoing, two tau candidate events were found from
4126 decay events in the data. A background of only 0.2 events is expected~\cite{opera}.

\section{Summary}\label{sec:sum}
 In this paper, the solar, atmospheric, and accelerator (except for T2K) neutrino results 
were discussed. From these data, the following neutrino oscillation parameters have been measured:
\begin{eqnarray}
  \Delta m_{21}^2 & = & 7.44^{+0.20}_{-0.19} \times 10^{-5}eV^2 \hspace{0.5cm} 32.7 < \theta_{12} < 34.3 ~deg.~(solar,KamL.)\\
  \left|\Delta m_{32}^2\right| & = & 2.66^{+0.15}_{-0.40} \times 10^{-3}eV^2 \hspace{0.5cm} 38.7 < \theta_{12} < 51.9 ~deg.~(SK atm.)\\
                 & = & 2.39^{+0.09}_{-0.10} \times 10^{-3}eV^2 \hspace{0.5cm} 36.8 < \theta_{12} < 53.2 ~deg.~(MINOS)
\end{eqnarray}
In addition several important results have been discussed:
\begin{itemize}
\item The first pep solar neutrino observation has been achieved by Borexino.
\item From the day-night asymmetry measured in SK there is some hint of the earth-matter effect.
\item Neutrino and antineutrino oscillations are consistent in the MINOS data.
\item There is some hint of $\nu_{\tau}$ appearance in the OPERA data.
\end{itemize}
%

%

\end{document}